\documentclass[
 reprint,
 superscriptaddress,
 amsmath,amssymb,
 aps,
pra,
showkeys
]{revtex4-2}

\usepackage{soul}
\usepackage{float}
\usepackage{multibib}
\usepackage{hyperref}
\usepackage{xcolor}
\hypersetup{
    colorlinks,
    linkcolor={blue!50!blue},
    citecolor={blue!50!blue},
    urlcolor={blue!80!black}
}
\usepackage{siunitx}
\usepackage{import}
\usepackage{placeins}
\usepackage{xcolor}
\usepackage{lmodern}
\usepackage[braket, qm]{qcircuit}
\usepackage{graphicx}
\usepackage{dcolumn}
\usepackage{bm}
\usepackage{comment}
\usepackage[english]{babel}
\usepackage{blindtext}
\usepackage{physics}  
\usepackage{multirow}
\usepackage{array}
\usepackage{enumerate}

\usepackage{ulem}
\definecolor{myblue}{rgb}{0.0, 0.75, 1.0}
\definecolor{lightpink}{rgb}{0.9, 0.4, 0.38}

\begin{document}
\title{Raman Signatures of Single Point Defects in Hexagonal Boron Nitride Quantum Emitters}  

\author{Chanaprom Cholsuk}
\email{chanaprom.cholsuk@tum.de}
\affiliation{Department of Computer Engineering, TUM School of Computation, Information and Technology, Technical University of Munich, 80333 Munich, Germany}
\affiliation{Munich Center for Quantum Science and Technology (MCQST), 80799 Munich, Germany}

\author{Asl{\i} \surname{\c{C}akan}}
\affiliation{Department of Computer Engineering, TUM School of Computation, Information and Technology, Technical University of Munich, 80333 Munich, Germany}
\affiliation{Munich Center for Quantum Science and Technology (MCQST), 80799 Munich, Germany}

\author{Volker Deckert}
\affiliation{Institute of Physical Chemistry, Friedrich-Schiller University, 07743 Jena, Germany}
\affiliation{Leibniz Institute of Photonic Technology, 07745 Jena, Germany}
\affiliation{Abbe Center of Photonics, Institute of Applied Physics, Friedrich Schiller University Jena, 07745 Jena, Germany}

\author{Sujin Suwanna}
\affiliation{Optical and Quantum Physics Laboratory, Department of Physics, Faculty of Science, Mahidol University, Bangkok 10400, Thailand}

\author{Tobias Vogl}%
\email{tobias.vogl@tum.de}
\affiliation{Department of Computer Engineering, TUM School of Computation, Information and Technology, Technical University of Munich, 80333 Munich, Germany}
\affiliation{Munich Center for Quantum Science and Technology (MCQST), 80799 Munich, Germany}
\affiliation{Abbe Center of Photonics, Institute of Applied Physics, Friedrich Schiller University Jena, 07745 Jena, Germany}

\date{\today}

\begin{abstract}
Point defects in solid-state quantum systems are vital for enabling single-photon emission at specific wavelengths, making their precise identification essential for advancing applications in quantum technologies. However, pinpointing the microscopic origins of these defects remains a challenge. In this work, we propose Raman spectroscopy as a robust strategy for defect identification. Using density functional theory, we systematically characterize the Raman signatures of 100 defects in hexagonal boron nitride (hBN) spanning periodic groups III to VI, encompassing around 30,000 phonon modes. Our findings reveal that the local atomic environment plays a pivotal role in shaping the Raman lineshape, enabling the narrowing of potential defect candidates. Furthermore, we demonstrate that Raman spectroscopy can differentiate defects based on their spin and charge states as well as strain-induced variations, implying the versatility of this approach. Therefore, this study not only provides a comprehensive theoretical database of Raman spectra for hBN defects but also establishes a novel experiment framework for using tip-enhanced Raman spectroscopy to identify point defects. More broadly, our approach offers a universal method for defect identification in any quantum materials.
\end{abstract}


\maketitle

Point defects in solid-state quantum systems serve as sources of single-photon emission in quantum emitters by inducing two-level defect states within the mid-gap region.~The emission wavelength of a quantum emitter depends on the localization position of the defect state. This phenomenon is particularly pronounced in low-dimensional hexagonal boron nitride (hBN) due to its intrinsic large bandgap ($\sim$6 eV), which facilitates the formation of a multitude of defects \cite{10.1038/nnano.2015.242, 10.3390/nano12142427,10.1021/acsphotonics.8b00127,10.1039/c9nr04269e,10.1021/acs.nanolett.6b01368,10.1002/adom.202402508}. However, the large number of possible defect types in hBN complicate defect identification. To address this, various experimental and computational methods have been developed to accurately determine the microscopic origins of these defects \cite{10.1088/2053-1583/ab8f61,10.3389/frqst.2022.1007756,10.1021/acs.jpclett.3c01475,10.1038/s41563-020-0619-6}.\\
\indent The most prominent identification strategy involves comparing the optical properties obtained from experiments with those predicted by first-principle calculations.~These properties include the zero-phonon line (ZPL), phonon sideband, photoluminescence (PL) spectra, excited-state lifetimes, and dipole orientations. Numerous studies have investigated extensive defect types using this approach such as C$_\text{B}$V$_\text{N}$ \cite{10.1088/2053-1583/ab8f61,10.3389/frqst.2022.1007756}, carbon-based defects up to tetramers \cite{10.1021/acs.jpca.0c07339}, C$_\text{B}$C$_\text{N}$ \cite{10.1063/1.5124153}, C$_\text{B}$C$_\text{N}$C$_\text{N}$ \cite{10.1103/PhysRevMaterials.6.L042201,10.1038/s41563-021-00979-4}, C$_\text{B}$C$_\text{N}$C$_\text{B}$ \cite{10.1038/s41563-021-00979-4}, C$_\text{B}$ \cite{10.1038/s41563-021-00979-4}, C$_\text{N}$ \cite{10.1038/s41563-021-00979-4}, C$_\text{B}$C$_\text{N}$C$_\text{B}$C$_\text{N}$ \cite{10.1063/5.0147560,10.1021/acsnano.3c08940,10.1021/acs.jpclett.3c01475,10.1021/acsanm.4c02722}, O$_\text{N}$V$_\text{B}^{-1}$ \cite{10.1021/acs.jpclett.2c02687}, N$_\text{B}$V$_\text{N}$ \cite{10.1038/s41534-020-00312-y}, and V$_\text{B}^{-1}$ \cite{10.1038/s41524-020-0305-x,10.1038/s41563-020-0619-6} etc. Despite its utility, such optical properties can still be prone to misassignment for several reasons \cite{10.1021/acs.jpclett.3c01475}.~(i) Many defects exhibit similar ZPL and PL lineshapes, making differentiation challenging. (ii) Optical properties are highly sensitive to the accuracy of excited-state computations. In particular, plane-wave density functional theory (DFT), while effective for ground-state calculations, faces limitations in describing excited states. (iii) Certain defects undergo symmetry changes between ground and excited states due to the Jahn-Teller effect, so this complicates ZPL calculations. For example, V$_\text{B}^{-1}$ transitions from $D_{3h}$ in the ground state to $C_{2v}$ in the excited state \cite{10.1038/s41524-020-0305-x}, while N$_\text{B}$V$_\text{N}$ transitions from $C_s$ in the ground state to $C_{2v}$ in the excited state \cite{10.1038/s41534-020-00312-y}. (iv) Lifetimes calculated via DFT often diverge from experimental values due to substrate-induced Purcell effects \cite{10.1021/acs.jpclett.3c01475}.\\
\indent To mitigate the uncertainties inherent in optical characterization, electron paramagnetic resonance (EPR) spectroscopy has been proposed as an alternative approach. This method was successfully employed to identify V$_\text{B}^{-1}$ by achieving consistency between experimental data and simulation \cite{10.1038/s41563-020-0619-6}. Subsequently, EPR spectroscopy has been extended to the identification of O$_\text{N}$V$_\text{B}^{-1}$ \cite{10.1021/acs.jpclett.2c02687} and C$_\text{B}$V$_\text{N}$ defects \cite{10.3389/frqst.2022.1007756}. The reliability of this method stems from its focus on the ground-state calculations, specifically the zero-field splitting (ZFS) and hyperfine interactions, which uniquely characterize the interactions between the defect and neighboring nuclei. Nevertheless, challenges remain in implementing EPR-based methods. (i) Plane-wave DFT often overestimates ZFS values in high-spin systems due to spatial separation and shape differences in spin-up and spin-down orbitals, leading to non-eigenstate descriptions. Thus, spin decontamination techniques are required to address this issue \cite{10.1103/PhysRevResearch.2.022024}. (ii) Computational errors in hyperfine interaction values can increase with the defect’s size and complexity, necessitating alternative methodologies for large defect complexes \cite{10.1038/s42005-024-01668-9}. (iii) EPR spectral peaks may appear similar if defects share analogous neighboring nuclei and hyperfine interactions. (iv) Defects with singlet-spin configurations present additional challenges for the EPR-based identification approach.\\
\indent Such limitations in current methods underscore the need for alternative approaches to achieve accurate defect identification. This paper, therefore, proposes leveraging the Raman signatures of point defects in hBN as another complementary strategy. Raman spectroscopy has the potential for this purpose, as it can capture the unique intrinsic vibrations of atoms, directly reflecting their microscopic environments \cite{10.1038/s41699-020-0140-4,10.1038/s41467-020-16529-6,10.1021/jacs.4c07812}. When a defect is introduced into a host material, it induces localized vibrational modes or perturbs existing phonon modes of the pristine structure. These perturbations manifest as shifts, broadening, or the emergence of new peaks in the Raman spectrum, making the defect-induced Raman feature a promising approach for uniquely characterizing point defects. Moreover, in theory, Raman simulation primarily involves ground-state calculation, thereby circumventing the challenges associated with excited-state DFT calculations.\\
\indent Thus far, the Raman spectroscopy has already been used to differentiate composition in materials \cite{10.1038/s41597-023-01988-5}, types of pristine two-dimensional materials \cite{10.1038/s41467-020-16529-6}, and ensembles of point defects in materials \cite{10.1021/jacs.4c07812,10.1088/1361-648X/ada106}. Despite its potential, identifying a single localized point defect remains challenging. Computationally, calculating the Raman spectrum for a defect still demands substantial resources, as it requires fully accounting for all phonon modes and their polarizability changes. Machine learning and deep learning methods have been employed to address this computational bottleneck \cite{10.1038/s41467-020-16529-6,10.1088/1361-648X/ada106,10.1021/acs.jpcc.4c00886,10.1021/jacs.4c07812}, but these approaches may fail to capture some Raman peaks due to the incomplete analysis of all phonons for certain defects. Experimentally, the spatial resolution of conventional Raman spectroscopy is likely insufficient to probe individual point defects \cite{10.1016/S0009-2614(99)01451-7,10.1039/B705967C,10.1021/nl300901a}. As a consequence, the Raman signatures of point defects, especially in hBN, have remained immensely uncharacterized both theoretically and experimentally.\\ 
\indent In this study, we have, therefore, performed intensive calculations of Raman signatures for 100 defects in hBN. The full consideration of all phonon modes has been made, which involved three spatial degrees of freedom of all atoms in a structure. This yields around 300 modes for each defect and, hence, around 30,000 modes for all studied defects. Initially, we benchmark our calculated Raman spectrum with the experimental Raman spectrum of pristine hBN. Subsequently, we demonstrate how defect-induced Raman shifts can serve as unique fingerprints for distinguishing various defects, focusing on point defects and complexes spanning periodic groups III to VI. As we simulate Raman signatures of defects in hBN to be verified in future experiments with tip-enhanced Raman spectroscopy, we develop classification metrics that consider not only the maximum Raman shift peaks but also the overall Raman spectral profiles. This approach ensures that defects with similar peak shift positions but distinct spectral profiles can be accurately distinguished. To achieve this, we employed the dynamic time warping (DTW) method. Finally, we propose an experimental concept relying on tip-enhanced Raman spectroscopy to enable the experimental realization of defect identification via Raman spectroscopy.\\
\indent As such, this work has developed a comprehensive theoretical database for a wide array of hBN defects, relying solely on the ground-state DFT calculations. This approach offers a robust and reliable foundation for assigning the microscopic origins of observed Raman signatures. Moreover, the methodology and findings presented here highlight the potential of Raman spectroscopy as a novel, versatile, and effective tool for defect identification, not only in hBN but also in other quantum materials. By bridging theoretical insights with experimental feasibility, this study paves the way for advancing defect characterization techniques.

\section*{Results}
The Raman spectra were systematically analyzed through the following four parts. First, the Raman spectra of pristine hBN, carbon-based defects, oxygen-based defects, and other defect types from periodic groups III to VI were characterized (see Supplementary S1 for all complete Raman spectra). Second, the similarity of the Raman lineshapes was quantitatively assessed. Third, the distribution of Raman shifts for 100 defects was demonstrated. Finally, an experimental proposal was developed and elaborated in this section.

\begin{figure*}[ht!]
    \centering
    \includegraphics[width=1\textwidth]{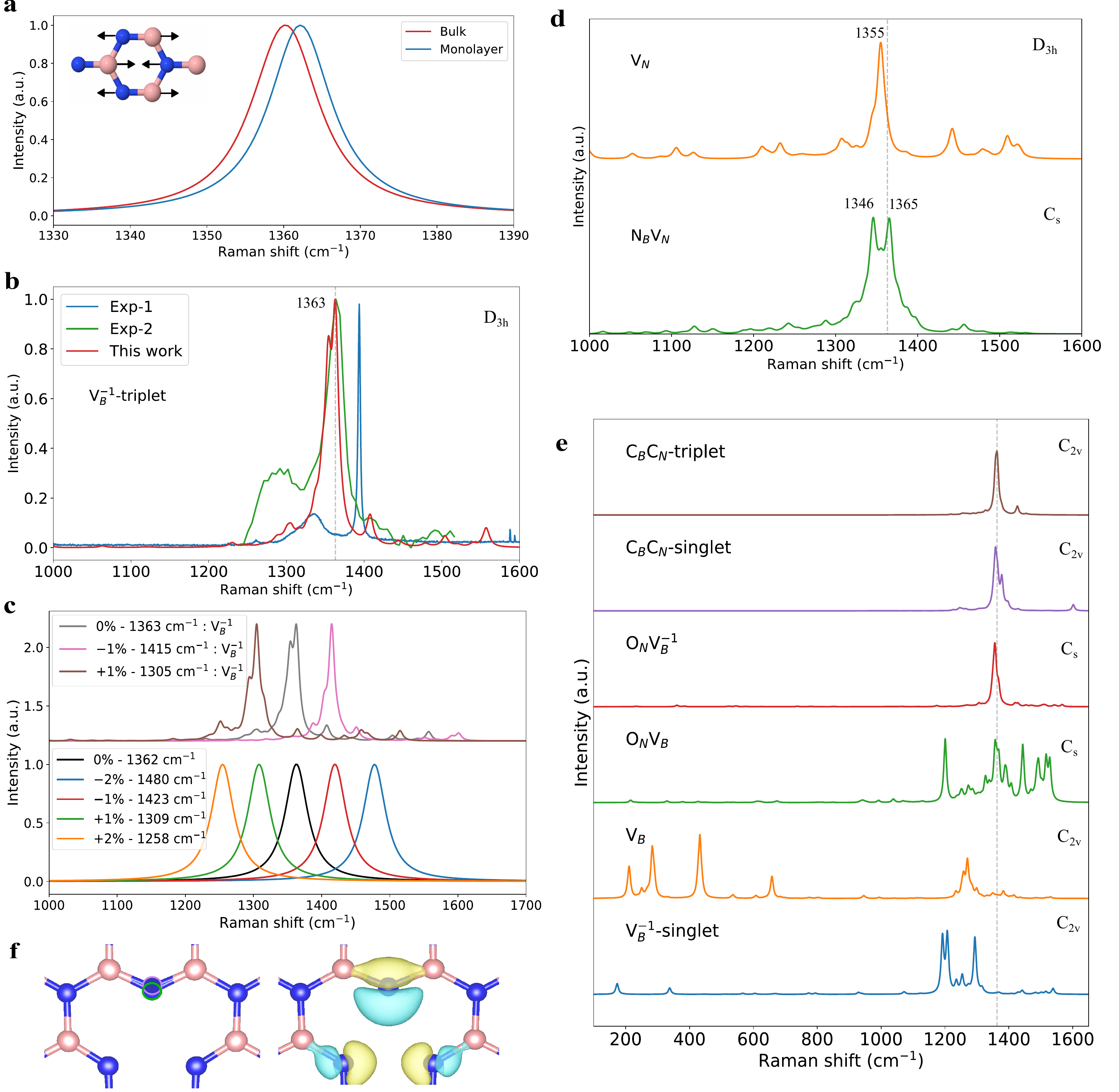}
    \caption{\textbf{Raman spectra and their evolution with the presence of defect and strain. a} Raman spectra of pristine hBN monolayer and bulk. The phonon mode in the inset is E$_{2g}$ mode, which is responsible for both 1362.2 cm$^{-1}$ in monolayer and 1361.1 cm$^{-1}$ in bulk, respectively. \textbf{b} Comparison of Raman signature in V$_\text{B}^{-1}$. The red curve is the simulated Raman spectrum in this work while the blue (Exp-1) and green (Exp-2) curves are experimental spectra from Refs.~\cite{10.1021/acs.chemmater.1c02849} and \cite{10.48550/arXiv.2501.18481}, respectively. \textbf{c} Simulated Raman spectra of defects with bi-axial strains where the indicated numbers refer to the strain value and the peak frequency, respectively. \textbf{d} Simulated Raman spectra of native defects. \textbf{e} Simulated Raman spectra of defects with different spin and charge states. The dashed gray line represents the Raman shift peak of the pristine hBN monolayer. The label in the right corner signifies the group symmetries. \textbf{f} (left) Geometry difference of V$_\text{B}^{-1}$ between triplet and singlet spin states where the pink and green circles represent an overlapping of a nitrogen atom. A nitrogen atom in the optimized V$_\text{B}^{-1}$ singlet structure is distorted by being pushed down towards the vacancy. (right) Charge density differences of V$_\text{B}^{-1}$ between triplet and singlet spin states as a result of geometry changes. The yellow and cyan isosurfaces represent the positive and negative charge density, respectively.  }
    \label{fig:hBNpristine}
\end{figure*}

\subsection*{Raman Signatures of Pristine hBN}
First, theoretical characterization of the Raman signature was conducted using a DFT calculation that considered full phonon modes and relied on the Raman assumption elaborated in the Methods section.
Fig.~\ref{fig:hBNpristine}(a) presents the Raman spectra of pristine hBN in both monolayer and bulk. For a 2$\times$2$\times$1 supercell, 24 and 49 phonon modes are considered for the monolayer and bulk, respectively. Among these, three correspond to acoustic phonons, while the rest are optical ones. The calculated Raman peaks yield 1362.2 cm$^{-1}$ and 1361.1 cm$^{-1}$, which are responsible by the E$_{2g}$ phonon mode depicted in the inset. These calculated values are in good agreement with the experimental values reported at 1367.5 cm$^{-1}$ for monolayer and 1365.1 cm$^{-1}$ for bulk, respectively \cite{10.1038/nnano.2015.242} and also consistent with other calculations \cite{10.1038/s41467-020-16529-6,10.1039/C6NR09312D}. We note that the small difference between simulation and experiment can be attributed to DFT functionals as studied earlier \cite{10.1039/C6NR09312D} or the local strain in the sample, as will be described later. As a large number of defects are considered in this work, the theoretical characterization of Raman spectra has been conducted based on the combination of PBE and HSE functionals to reduce computational cost (See Methods section and Supplementary Sec.~S2 for more details). When comparing monolayer and bulk structures, only slight differences in Raman shifts are observed in this study, agreeing well with experimental result \cite{10.1038/nnano.2015.242} and prior DFT calculation \cite{10.1039/C6NR09312D}. While most experiments create defects within bulk hBN, the defect-induced Raman signatures in this work have been analyzed exclusively in hBN monolayers. This choice is justified by the fact that bulk hBN exhibits a significantly larger number of phonon modes, which demands excessive computational resources. More importantly, experiments utilizing tip-enhanced Raman spectroscopy are generally limited to detecting vibrations in the topmost layers. As such, the theoretical characterization of hBN monolayers presented here remains directly relevant to experiments performed on both monolayer and bulk structures.

\subsection*{Raman Signatures of Native hBN Defects and Effects of External Factors}
In this section, native defects are introduced in the hBN monolayer, which has a supercell size of $7\times7\times1$ with a 15~\AA~vacuum layer. See the Methods section for the computational details and Supplementary S3 for the convergence test.~Fig.~\ref{fig:hBNpristine}(b) illustrates how a defect modifies the Raman lineshape, introduces new peaks, and shifts existing peaks. Here, we start with the well-known V$_\text{B}^{-1}$ defect. Its symmetry is classified as the $D_{3h}$ point group, characterized by three nitrogen atoms symmetrically arranged around the vacancy, forming bonds that preserve this symmetry. Based on the 7$\times$7$\times$1 supercell, 294 vibrational modes are considered in this defect. We found that five of them can be identified as acoustic phonons while the rest are optical phonons. The Raman spectrum, as shown in Fig.~\ref{fig:hBNpristine}(b), indicates a slight peak shift from the pristine hBN to 1363 cm$^{-1}$, accompanied by the appearance of additional peaks, especially at 1304 cm$^{-1}$ and 1407 cm$^{-1}$. The most pronounced Raman shift peak at 1363 cm$^{-1}$ can be attributed to changes in vibrational modes caused by the vacancy. Unlike the pristine hBN, where horizontal stretching dominates (as shown in the inset of Fig.~\ref{fig:hBNpristine}(a)), the V$_\text{B}^{-1}$ defect exhibits slight vertical stretching while still remaining horizontal stretching. As this phonon mode is found to still be an in-plane vibration similar to the E$_{2g}$, this results in a slightly shifted Raman peak from the pristine hBN. This is in good agreement with experimental observations where the most dominant Raman shift peak remains close to the pristine hBN \cite{10.1021/acs.chemmater.1c02849, 10.48550/arXiv.2501.18481}. For our two additional peaks at 1304 cm$^{-1}$ and 1407 cm$^{-1}$, the former peak corresponds to the experimental peaks at 1295 cm$^{-1}$ by Ref.~\cite{10.48550/arXiv.2501.18481} and at 1335 cm$^{-1}$ by Ref.~\cite{10.1021/acs.chemmater.1c02849}, while the latter peak corresponds to the small experimental bump at 1400 cm$^{-1}$ by Ref.~\cite{10.48550/arXiv.2501.18481}. We found that the intensity of our simulated Raman shift peak at 1304 cm$^{-1}$ is consistent with that of Exp-1 \cite{10.1021/acs.chemmater.1c02849} with a shift by 30 cm$^{-1}$, but is lower than that of Exp-2 \cite{10.48550/arXiv.2501.18481} without any shift in the wavenumber. The intensity discrepancy can be explained by the difference in V$_\text{B}^{-1}$ concentration. In the simulation, the Raman spectrum has been characterized based on the single point defect, that is isolated from other neighboring defects, while, in the experiment, the Raman measurement has been carried out in an ensemble of defects. We believe that the V$_\text{B}^{-1}$ concentration in Ref.~\cite{10.1021/acs.chemmater.1c02849} might be lower than that in Ref.~\cite{10.48550/arXiv.2501.18481}, making it more comparable to our isolated defect calculations. While simulating higher concentrations of V$_\text{B}^{-1}$ (a defect ensemble) can likely broaden the peak, as observed by another machine learning model \cite{10.1088/1361-648X/ada106}, such the simulation using DFT alone would be computationally prohibitive. In addition, for the shift of the entire Exp-1 spectrum compared to the simulated and Exp-2 spectra, we speculate that this stems from the local strain effect in the sample. As shown in Fig.~\ref{fig:hBNpristine}(c), strain can shift the entire Raman characteristics while preserving its lineshape for both pristine hBN and V$_\text{B}^{-1}$ defect. Specifically, only the peak positions are shifted. We found that the Exp-1 spectrum is shifted by 30 cm$^{-1}$ from the simulated spectrum. While this number corresponds to approximately 0.5\% tensile strain, as interpolated from Fig.~\ref{fig:hBNpristine}(c), we note that further investigation should be carried out. Overall, the behavior of the spectrum shift suggests that strain-induced modifications directly alter the vibrational frequencies while maintaining the changes in polarizability. 
We note that the strain range studied in this work assumes a typical range that we extracted from previous measurements using second-harmonic generation (SHG) \cite{10.1021/acsnano.3c08940}. See the Methods section for further details.\\
\indent When compared to the V$_\text{N}$ defect in Fig.~\ref{fig:hBNpristine}(d), the Raman lineshape of V$_\text{N}$ appears similar to V$_\text{B}^{-1}$, with only a slight peak shift. This similarity can be explained by the similar local environment, which consists of three neighboring atoms around a vacancy. This results in the same $D_{3h}$ symmetry; hence, leads to similar phonon vibrational modes and polarizability.\\
\indent In contrast, the Raman spectrum for the N$_\text{B}$V$_\text{N}$ defect is different. This is due to different local environments given that after the geometry optimization, the N$_\text{B}$ atom shifts downward (out-of-plane), resulting in a $C_s$ symmetry group. Furthermore, two split peaks have been revealed, with prominent ones at 1346 cm$^{-1}$ and 1365 cm$^{-1}$, as illustrated in Fig.~\ref{fig:hBNpristine}(d). This splitting arises from slight variations in the phonon modes.\\
\indent Returning to the Raman signature of V$_\text{B}^{-1}$, this defect is known to exhibit intersystem crossing, which enables the transition from triplet to singlet spin configurations \cite{10.1038/s41524-020-0305-x}. The question arises whether the Raman signature of its singlet state results in different characteristics. As depicted in Fig.~\ref{fig:hBNpristine}(e), the Raman signature of V$_\text{B}^{-1}$ in the singlet state is indeed distinct from its triplet counterpart. This difference arises from variations in their optimized geometries. Specifically, as shown in Fig.~\ref{fig:hBNpristine}(f), the position of a neighboring nitrogen atom differs between the two spin configurations, leading to variations in charge density and, consequently, polarizability. Then, the neutral V$_\text{B}$ defect has also been investigated. Fig.~\ref{fig:hBNpristine}(e) illustrates that the Raman signature is also uniquely different from others. To verify the robustness of these observations, we extended our study to oxygen and carbon defects, as demonstrated in Fig.~\ref{fig:hBNpristine}(e). We found that the ability to distinguish charge states and spin configurations via Raman spectroscopy remains valid. These results emphasize the capability of Raman spectroscopy to effectively differentiate between charge states and spin configurations of point defects.\\

\subsection*{Raman Signatures of Carbon-based and Oxygen-based hBN Defects}
\begin{figure*}[ht!]
    \centering
    \includegraphics[width=1\textwidth]{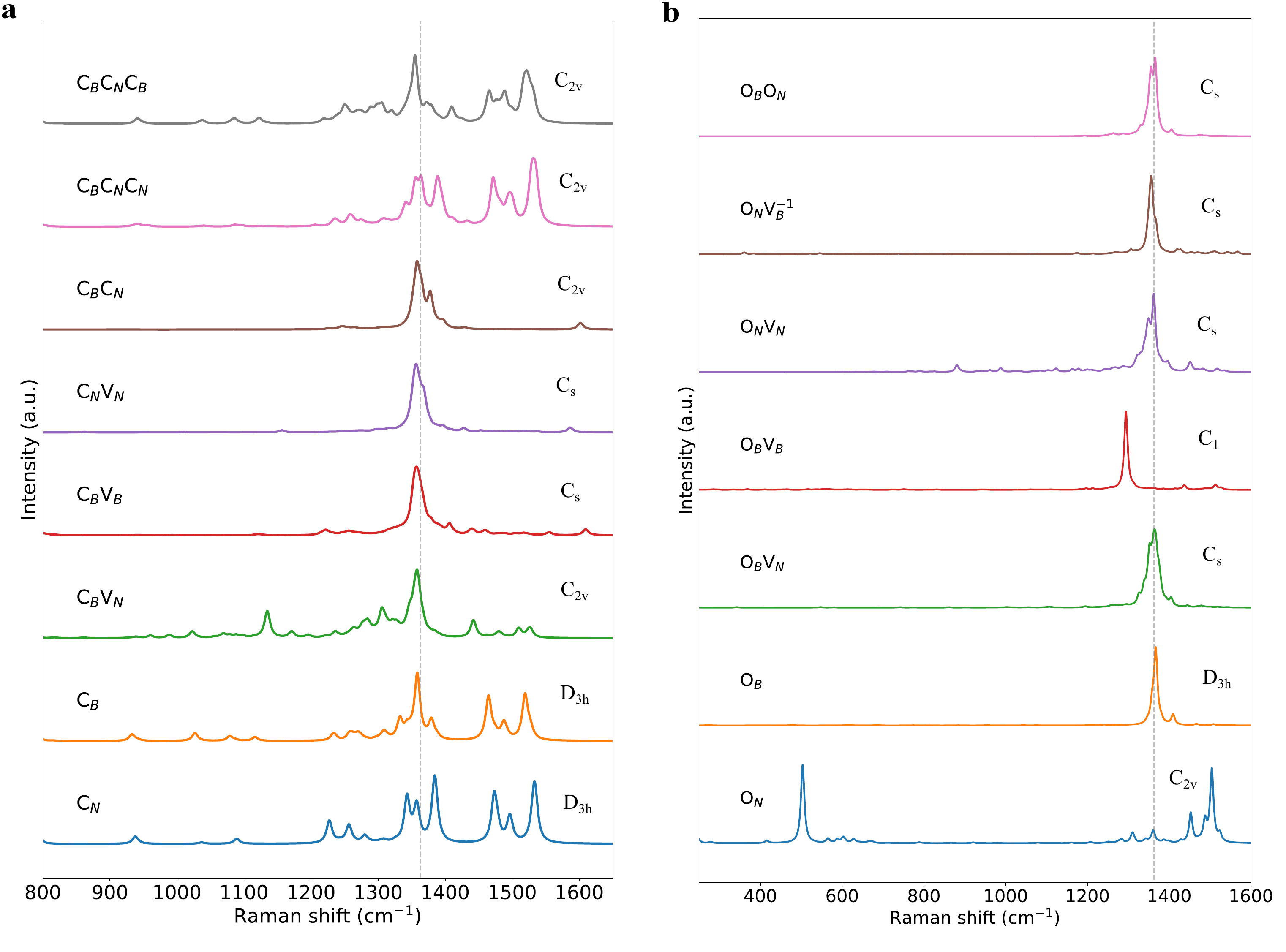}
    \caption{\textbf{Simulated Raman spectra of C-based and O-based defects. a} C-based defects. \textbf{b} O-based defects. The dashed gray line represents the Raman shift peak of the pristine hBN monolayer. The label in the right corner signifies the inherent group symmetries.}
    \label{fig:C-O-defects}
\end{figure*}

Having considered the effect of native defects, spin states, and strain on the Raman spectra, we now expand our characterization to the Raman spectra of C-based and O-based defects. In this section, certain defects, which were considered as promising candidates for experiments, are demonstrated \cite{10.1088/2053-1583/ab8f61,10.3389/frqst.2022.1007756, 10.1063/1.5124153,10.1103/PhysRevMaterials.6.L042201,10.1038/s41563-021-00979-4,10.1038/s41563-021-00979-4,10.1021/acs.jpclett.2c02687}. Other defects can be found in Supplementary S1. It is important to note that all Raman spectra in this study are normalized to their maximum peak intensities. Although peak intensities can vary between different defects and may be used for comparison, our analysis shows that normalization does not alter the relative spectral features. Therefore, for consistency and clarity, all Raman intensities presented in this study are normalized.\\
\indent As presented in Fig.~\ref{fig:C-O-defects}(a), the dominant peak for most carbon defects remains near the Raman shift of pristine hBN. However, the overall lineshapes vary significantly and can be categorized based on the local environment. For instance, the C$_\text{N}$ and C$_\text{B}$ defects exhibit similar lineshapes due to their comparable local environments and the same $D_{3h}$ symmetry group. Notably, despite belonging to the same symmetry group as the V$_\text{B}^{-1}$ and V$_\text{N}$ defects, carbon substitution modifies the polarizability, resulting in distinct Raman spectra.\\
\indent In the presence of a vacancy, the Raman spectra of C$_\text{B}$V$_\text{B}$ and C$_\text{N}$V$_\text{N}$ defects appear similar, whereas that of C$_\text{B}$V$_\text{N}$ differs. This disparity arises from differences in local environments, particularly as both C$_\text{B}$V$_\text{B}$ and C$_\text{N}$V$_\text{N}$ have the $C_s$ symmetry group, whereas C$_\text{B}$V$_\text{N}$ is classified as the $C_{2v}$ group. Consequently, variations in phonon modes and polarizability lead to distinct spectral features.\\
\indent Then, considering the dimer and trimer defects, the Raman spectra are unique and depend on the number of carbons. Although all C$_\text{B}$C$_\text{N}$, C$_\text{B}$C$_\text{N}$C$_\text{N}$, and C$_\text{B}$C$_\text{N}$C$_\text{B}$ defects are in the $C_{2v}$ group, only C$_\text{B}$C$_\text{N}$C$_\text{N}$ and C$_\text{B}$C$_\text{N}$C$_\text{B}$ exhibit similar Raman lineshapes. This divergence is attributable to differences in local environments.\\
\indent Turning to consider O-based defects, Fig.~\ref{fig:C-O-defects}(b) reveals behavior analogous to that observed for C-based defects. Specifically, most defects exhibit a prominent Raman shift near that of pristine hBN, with the exceptions of  O$_\text{N}$  and  O$_\text{B}$V$_\text{B}$. For O$_\text{B}$ and O$_\text{N}$, while their local environments appear similar, their Raman signatures are different. This is due to the difference in optimized geometry where O$_\text{N}$ belongs to the $C_{2v}$ symmetry group, whereas O$_\text{B}$  is classified as $D_{3h}$. This distinction, in turn, affects polarizability and, consequently, the Raman spectrum. Then, comparing O$_\text{B}$ and C$_\text{B}$ as well as O$_\text{N}$ and C$_\text{N}$, even though these defect pairs have similar local environments, their Raman lineshapes are distinct. This can be attributed to the dopants where C and O influence the polarizability differently. Then, for the O$_\text{B}$V$_\text{N}$ and O$_\text{B}$V$_\text{B}$, the Raman peaks are different due to the different symmetries and polarizability. The same behavior and justification are also applied for O$_\text{N}$V$_\text{N}$, O$_\text{N}$V$_B^{-1}$, and O$_\text{B}$O$_\text{N}$ defects.

\subsection*{Similarity of Raman lineshape based on Local Environment}
As suggested in the previous section, a similar local environment tends to produce similar Raman lineshapes. In this section, we further investigate whether identical local environments, but arising from different periodic groups, result in similar Raman lineshapes and peah shifts. The similarity is quantified using the DTW distance with the computational details described in the Methods section. To facilitate this analysis, each subsection is organized based on the corresponding local environment.

\begin{figure*}[ht!]
    \centering
    \includegraphics[width=1\textwidth]{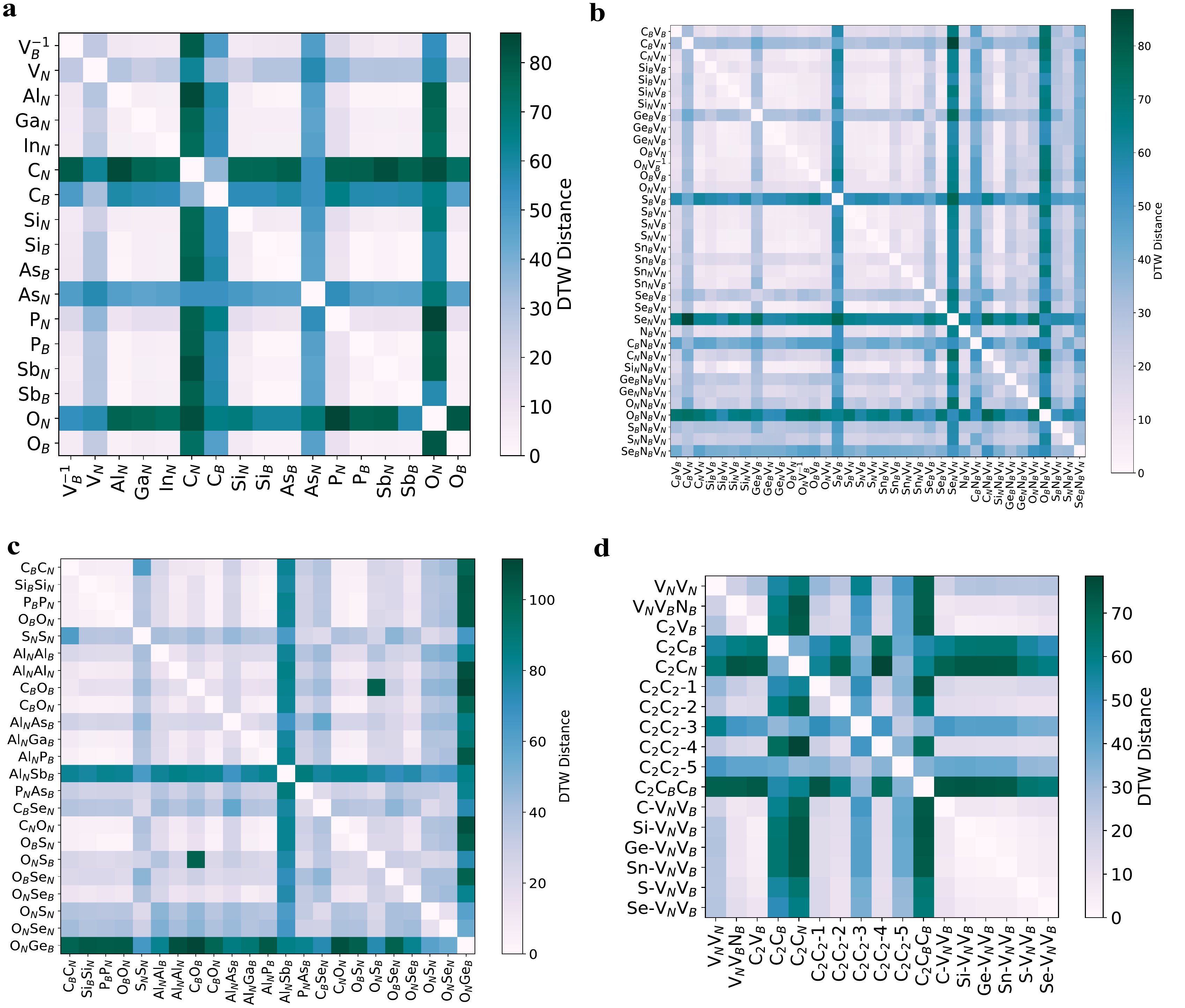}
    \caption{\textbf{Similarity of Raman signatures of hBN defects quantified by DTW distance.} \textbf{a} Point defects without any vacancy. \textbf{b} Point defects with one vacancy. \textbf{c} Two adjacent substitutional defects without vacancy. \textbf{d} Defect complexes where C$_2$ stands for C$_\text{B}$C$_\text{N}$. The C$_2$C$_2$-number signifies the structural configuration, which is displayed in Supplementary S2.}
    \label{fig:dtw_dist}
\end{figure*}

\subsubsection*{Point defects}
In the simplest defect type, a defect is a substitution of an impurity. This makes the local environment one impurity at the center surrounded by three host atoms. Fig.~\ref{fig:dtw_dist}(a) illustrates the similarity of Raman lineshapes and peaks for such defect types across periodic groups. The results indicate that most point defects exhibit comparable Raman features, as reflected by the low DTW distance, irrespective of the periodic group. Furthermore, the spectrum of V$_\text{B}^{-1}$ defect is also similar to other antisite defects. However, exceptions are observed for V$_\text{N}$, C$_\text{N}$, C$_\text{B}$, and O$_\text{N}$, where the Raman lineshapes differ due to large variations in the number of dominant peaks (see Supplementary S1 for the complete spectra). This finding suggests that defects containing any of these four defects can be likely identified using Raman spectroscopy, while other defects can use Raman spectra as a complementary technique.

\subsubsection*{Substitutional defects together with one vacancy}
The next question is whether the Raman spectrum can effectively distinguish a defect consisting of a vacancy adjacent to a substitutional defect. As shown in Fig.~\ref{fig:dtw_dist}(b), most defects seem identical, as seen from the low DTW distance. When comparing X$_\text{B}$V$_\text{N}$ and X$_\text{N}$V$_\text{B}$, the results indicate that their lineshapes are nearly identical when X represents the same atom, such as in Si$_\text{B}$V$_\text{N}$ and Si$_\text{N}$V$_\text{B}$. However, configurations like X$_\text{N}$V$_\text{N}$ and X$_\text{B}$V$_\text{B}$ can yield distinct Raman spectra due to differences in local environments. \\
\indent When comparing across periodic groups such as Si$_\text{B}$V$_\text{N}$ and Ge$_\text{B}$V$_\text{N}$, Fig.~\ref{fig:dtw_dist}(b) highlights a low DTW distance, indicating that an impurity affects the Raman spectrum similarly, regardless of its periodic group. However, exceptions exist, particularly for defects like C$_\text{B}$V$_\text{N}$, Ge$_\text{B}$V$_\text{B}$, S$_\text{B}$V$_\text{B}$, Se$_\text{B}$V$_\text{B}$, and Se$_\text{N}$V$_\text{N}$ defects, where the differences are more pronounced.\\
\indent Considering N$_\text{B}$V$_\text{N}$-related defects, most exhibit similar signatures, characterized by two distinct split peaks, closely resembling those of the N$_\text{B}$V$_\text{N}$ defect, as shown in Fig.~\ref{fig:hBNpristine}(d). See Supplementary S1 for the complete Raman spectra.  This highlights the uniqueness of this defect type and its potential for identification based on Raman spectral signatures.

\subsubsection*{Two substitutional defects without vacancy}
We now turn to examine two adjacent substitutional defects without a vacancy. As depicted in Fig.~\ref{fig:dtw_dist}(c), Si$_\text{B}$Si$_\text{N}$, P$_\text{B}$P$_\text{N}$, O$_\text{B}$O$_\text{N}$, C$_\text{B}$O$_\text{N}$, Al$_\text{N}$P$_\text{B}$, and O$_\text{B}$S$_\text{N}$ tend to exhibit similar lineshapes among each other. Upon further analysis, these defects were found to share the same $C_s$ point group symmetry, indicating that their comparable lineshapes are likely a result of experiencing similar local atomic environments. However, there is more to the local environment than group symmetry. For example, Al$_\text{B}$Al$_\text{N}$, whose symmetry also belongs to the $C_s$ group, displays a different Raman lineshape compared to the aforementioned defects. Furthermore, when taking into account other defect types, it becomes challenging to draw a conclusive relationship, as the Raman lineshapes of some defects are similar while others vary significantly between different defects. This observation suggests that, in this defect type, the lineshape is predominantly governed by the type of impurity and the local atomic environment surrounding the defect. As a result, considering individual lineshapes for each defect becomes indispensable.

\subsubsection*{Defect complexes}
Finally, we investigate the applicability of Raman spectroscopy for identifying defect complexes. As illustrated in Fig.~\ref{fig:dtw_dist}(d), the results clearly demonstrate that the local environment plays a pivotal role in determining the similarity of Raman lineshapes. Specifically, the defects can be categorized into four distinct groups. First, V$_\text{N}$V$_\text{N}$ has the unique lineshape from others due to different local environments. Second, bivacancy-related defects (X-V$_\text{B}$V$_\text{N}$) seem to have similar lineshapes regardless of the single impurity type, implying that the bivacancy structure dominates the vibrational characteristics. Third, trimers of C-based defects are likely to have comparatively similar lineshapes than others. Last, C$_2$C$_2$-1, C$_2$C$_2$-2, and C$_2$C$_2$-4 defects seem to have comparatively similar lineshapes than those of C$_2$C$_2$-3 and C$_2$C$_2$-5. This can be described by the differences in geometries of C$_2$C$_2$, as shown in Supplementary S3. The lineshape of C$_2$C$_2$-1 is the most unique from others as all carbons are closely attached. Meanwhile, the lineshapes of C$_2$C$_2$-2 and C$_2$C$_2$-4 defects are similar as both defects have a similar arrangement of one carbon substitution far apart from the connected three carbon substitutions. Similarly, C$_2$C$_2$-3 and C$_2$C$_2$-5 have the same defect arrangement by separating one carbon substitution from the edge of the three connected carbons, so this results in similar Raman lineshapes.

\subsection*{Distribution of Raman Shift Peaks in hBN Defects}
\begin{figure*}[ht!]
    \centering
    \includegraphics[width=1\textwidth]{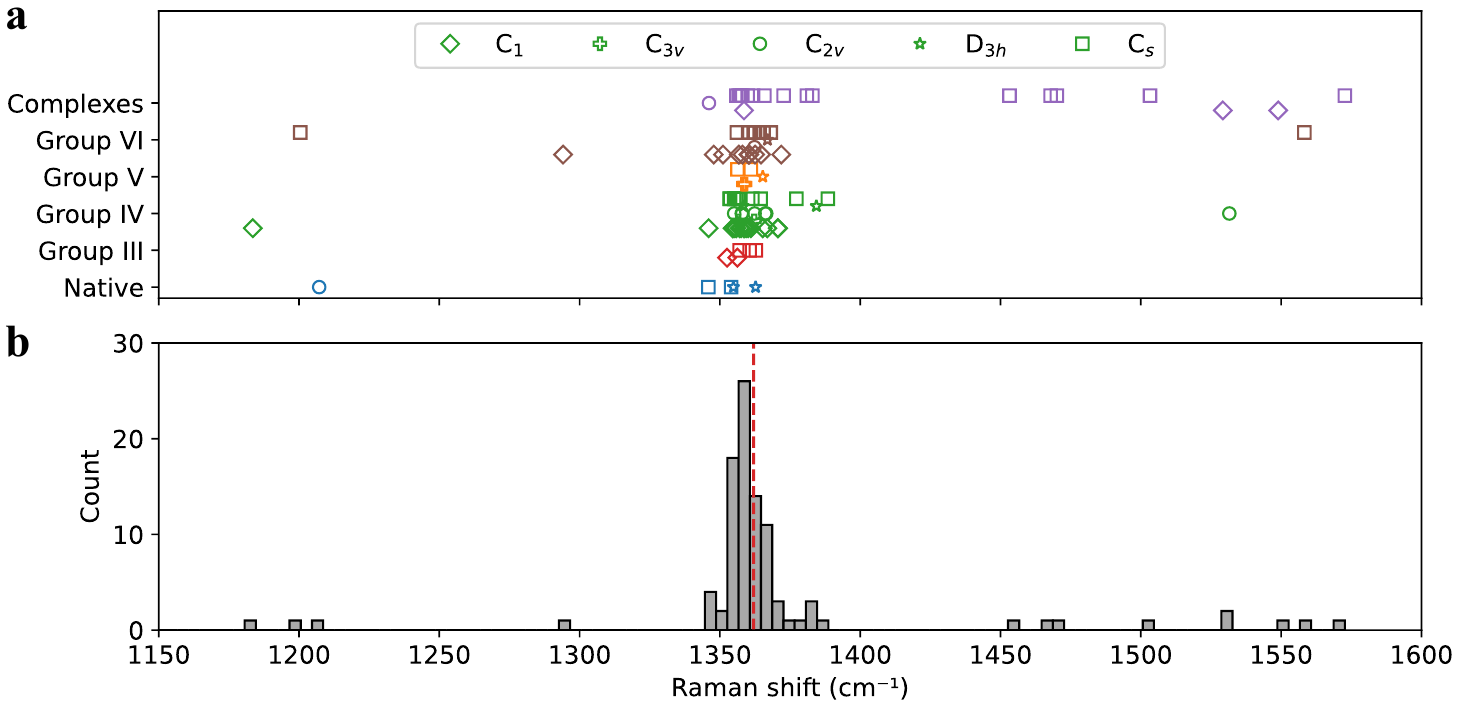}
    \caption{\textbf{Distribution of Raman shift peaks in 100 hBN defects.} \textbf{a} Distribution is classified by periodic groups, green, brown, and purple represent where blue, red native, group-III, group-IV, group-V, group-VI, and complex defects, respectively. The symbols indicate the corresponding group symmetries. \textbf{b} Histogram of the most dominant Raman shift peaks with a bin width of 4 cm$^{-1}$. The red dashed line marks the pristine hBN Raman shift at 1362.2 cm$^{-1}$. Note that three defects fall outside the displayed range: O$_\text{N}$ with a peak at 503.7 cm$^{-1}$, V$_\text{B}^{0}$ with a peak at 432.5 cm$^{-1}$, and As$_\text{N}$ with a peak at 2332.0 cm$^{-1}$.}
    \label{fig:distribution}
    
\end{figure*}

As mentioned in the previous section, the Raman lineshapes are primarily influenced by the local environment rather than the group symmetries. In this section, we now examine the distribution of the most prominent Raman shift peaks in various hBN defects. Fig.~\ref{fig:distribution}(a) reveals no clear correlation between the periodic group or symmetry group of a defect and its Raman shift peak position. In addition, as shown in the histogram in Fig.~\ref{fig:distribution}(b), the majority of Raman shift peaks cluster near the pristine hBN peak, particularly within the range of $\pm$ 8 cm$^{-1}$ from 1362.2 cm$^{-1}$. This suggests that, in terms of the highest Raman peak position, despite the diversity of defect types, many defects induce only slight variations from the host hBN.

\subsection*{Proposal for a Suitable Experiment}
While the  Raman spectra of the point defects in hBN can be well predicted, a major challenge from an experimental point of view is the local distinction of single point defects and the reliable discrimination of the strong background signal of the defect-free volume, respective area.  A standard confocal Raman microscopy approach, even when using liquid immersion lenses to improve the numerical aperture, cannot go far beyond an illumination spot of 200 nm in diameter when using visible excitation wavelengths. Finding a single defect, even in the best-case scenario of a single monolayer, is like looking for a needle in a haystack. To illustrate, if a defect is confined to a 1 nm region (which is likely overestimated), the 200 nm diffraction-limited spot encompasses a volume containing over 35,000 times more atoms than the 1 nm region. Considering that this 1 nm estimate for a single defect may already be an overestimation, this disparity further reduces the relative intensity of the defect signal against the background. Moreover, with a typical 16-bit dynamic range of charge-coupled device (CCD) cameras (2$^{16}$ $\approx$ 65,000), even a monolayer sample approaches the theoretical detection limit, making accurate measurements exceptionally challenging. All these atoms contribute to the pristine Raman peak, while only a few atoms contribute to the defect Raman signature. The Raman signal is therefore strongly influenced by the pristine hBN.\\
\indent Ideally, the problem can be addressed by a local discrimination, for instance by using near-field optical microscopy and its adaption to Raman spectroscopy, namely tip-enhanced Raman scattering (TERS).  TERS has been successfully adapted to 2D materials \cite{10.1038/s43586-024-00323-5} and particularly the sub-nanometer resolution capability. Already edge transitions/defects in MoS$_2$ have been documented using TERS, clearly indicating that a defect assessment is possible \cite{10.1038/s41467-019-13486-7}. Still a major challenge is the localization of a defect. Relying entirely on TERS alone is in most cases too time consuming. Assuming an optimistic acquisition time of 100 ms/spectrum is sufficient to find a defect, locating this defect would require at least a step size between TERS spectra of 1 nm. Mapping the aforementioned area of 0.2 $\times$ 0.2 $\mu$m would require more than one hour, which is a challenge for keeping the sample drift constant and relocating the probe back to the point defect site.  A potentially better solution would be a combination of different near-field optical techniques. An initial scattering-type Scanning Near-field Optical Microscope (sSNOM) experiment using a fast single-channel detector could probe the same surface area at least one order of magnitude faster. After such an experiment, the tip can be relocated to potential defect sites, and TERS scans on a much smaller area can be performed, which in turn either reduces the overall experimental time or allows for longer acquisition time per position to improve the signal-to-noise ratio. A similar approach would be the direct usage of stimulated Raman scattering and a plasmonic tip \cite{10.1021/acs.chemrev.6b00545,
10.1021/acs.accounts.5b00327}. Here either only a  single channel detector is required or the tip itself can act as a sensor via induced forces when the sample is resonantly enhanced. The distinction  of the actual Raman transition is based on the combination of laser lines and their respective tuning. To obtain a full spectrum at every site consequently requires the acquisition of as many images as required for a proper distinction or a full wavelength scan at  selected sites. The latter is most likely the fastest way. 
In summary, a (sub-)nanoscale distinction of single point defects via nanoscale Raman spectroscopy is feasible; however, at this time, at least a combination with single channel near-field optical techniques is desirable.

\section*{Discussion}
This study has proposed Raman signatures as an alternative and potentially effective strategy for defect identification.~Through a comprehensive theoretical analysis, Raman spectra of 100 defects hosted by hBN from periodic groups III to VI covering all phonon modes have been characterized. Importantly, we have demonstrated that these Raman signatures can be reliably predicted using the ground-state DFT calculations, avoiding the complexities associated with the calculation of excited-state properties. For pristine hBN, the Raman signature is dominated by the E$_{2g}$ mode, appearing near 1362 cm$^{-1}$. However, the presence of defects alters this spectrum, inducing peak shifts and the emergence of new peaks.\\
\indent In contrast to the pristine structure, where Raman spectrum can be systematically derived from crystallographic space groups and atomic positions using group theory \cite{10.1038/s41467-020-16529-6}, this framework becomes insufficient in the presence of defects.~Defect-induced local symmetry breaking introduces new vibrational modes and modifies polarizability, which manifests as distinct features in the Raman spectra. We found that the local environment plays more critical roles in the Raman lineshape than does group symmetry. That is, impurities originating from different periodic groups tend to produce similar Raman lineshapes if the local environment is similar. Conversely, defects with the same group symmetries may exhibit distinct Raman lineshapes and peaks, underscoring the necessity of theoretical characterization for each individual defect. This dependence of Raman signatures on local environments can be understood through Eqs.~\ref{eq:raman_act} and \ref{eq:dielectric}, where the Raman activity is determined by the derivative of the polarizability tensor with respect to atomic displacements. If defects are surrounded by similar arrangements of neighboring atoms (local environments), their charge density redistributions likely respond in a similar manner, leading to similar variations in the polarizability and, consequently, similar Raman spectral features. \\
\indent Additionally, this study has investigated the impact of external factors on the Raman signature. The results have shown that the Raman signature can capture the effects of different spin and charge states of the same defect species. These influences are reflected through the variation of the Raman lineshape, which can be explained by the differences in the optimized structures, which lead to changes in polarizability and, consequently, distinct Raman spectra. Furthermore, the presence of strain does not alter the lineshape but instead linearly shifts the Raman peak position. These results indicate the potential of Raman spectroscopy to distinguish these factors.\\
\indent Finally, we have proposed an experimental design for tip-enhanced Raman spectroscopy, which can potentially be applied in defect identification in both 2D and 3D materials, e.g., thin films, as long as the defect stays in the near-field range for TERS. Overall, the proposed Raman-based approach not only complements existing identification methodologies but also overcomes the current limitations in identification based on optical signatures. This offers a robust and versatile technique for advancing the understanding of point defects in quantum materials, not only in hBN but also in transition metal dichalcogenides and 3D materials such as diamonds. Additionally, our simulated Raman spectra can serve as a valuable benchmark for future theoretical modeling and experimental validation. 

\section*{Methods} \label{sec:method}
\subsection*{First-principle Calculation of Structural Optimization}
All DFT calculations were carried out using the Vienna Ab initio Simulation Package (VASP) \cite{vasp1,vasp2} based on the plane wave basis set and the projector augmented wave (PAW) pseudopotentials \cite{paw,paw2}. The pristine hBN bulk and monolayer structures were modeled in a 2$\times$2$\times$1 supercell. For the bulk, van der Waals forces were included using the DFT-D3 method of Grimme for dispersion correction \cite{10.1063/1.3382344, 10.1002/jcc.21759}, whereas for the monolayer, a vacuum region of 15 \AA~was introduced to eliminate interlayer interactions. As this study considered all phonon modes in the structure, only the monolayer was considered for point defect structures to reduce computational costs since the bulk introduces a significantly higher number of phonon modes. To model a defect, a 7$\times$7$\times$1 supercell was employed to mitigate defect interactions between neighboring cells. The structural optimizations were guided by our previously developed hBN defect database \cite{10.1021/acs.jpcc.4c03404}, with a force convergence threshold of 10$^{-2}$ eV/\AA, a total energy threshold of 10$^{-4}$ eV, and the use of the single $\Gamma$-point scheme. The database also facilitates total energy comparisons among singlet, doublet, and triplet spin configurations \cite{10.1002/adom.202302760}, ensuring that only the most energetically favorable configurations were selected for subsequent phonon and Raman calculations. For structural optimization, the screened hybrid functional of Heyd, Scuseria, and Ernzerhof (HSE06) was applied. While the screened Fock exchange ($\alpha$) ratio is sometimes tuned between 0.25 and 0.40 in other DFT studies to align with experimental optical band gaps \cite{10.1021/acs.jpclett.2c02687,10.1063/1.5124153,10.1103/PhysRevB.97.214104,10.1088/2053-1583/ab8f61}, our findings indicate that varying the $\alpha$ ratio during structural optimization has no significant impact on the Raman lineshape (see Supplementary S3). Consequently, the standard $\alpha$ ratio of 0.25 was consistently employed throughout this study.

\subsection*{First-principle Calculation of Vibrational Modes}
The vibrational modes were analyzed using the frozen phonon method at the $\Gamma$ point within the harmonic approximation. The VASP Transition State Tools (VTST) were utilized to generate displacements for all degrees of freedom, amounting to 3$N$, where $N$ represents the number of atoms in the structure. To ensure accurate force differences (curvature), the force convergence threshold and total energy convergence were set to 10$^{-10}$ eV/\AA~and 10$^{-8}$ eV, respectively. The VTST codes were further employed to calculate the dynamical matrix. Given the scale of this study, which analyzed 100 defect structures, approximately 30,000 phonon modes were considered in total. To minimize computational requirements, the PBE functional was employed to identify eigenfrequencies ($\nu_p$) and eigenvectors ($\vec{e}_p$) of each phonon mode ($p$) under a 3 $\times$ 3 $\times$ 1 reciprocal space grid. As the accuracy of phonon modes is highly sensitive to supercell size, a convergence test was performed. It was determined that a 7$\times$7$\times$1 supercell is sufficiently large to ensure the Raman spectrum remains unchanged. Detailed results of the convergence tests are provided in Supplementary S4.

\subsection*{Calculation of Raman Spectra}
The Raman spectrum was characterized under the common assumption of a plane-polarized incident laser beam with the beam’s direction, its polarization, and the observation direction all mutually perpendicular. This leads to the derivative of a Raman cross section for the Stokes component of Raman scattering from the $p$-th phonon mode provided by \cite{10.1007/3-540-11380-0_14,10.1103/PhysRevB.54.7830}
\begin{equation}
\frac{d \sigma_p}{d \Omega}=\frac{\left(2 \pi \nu_S\right)^4 h}{8 \pi^2 \nu_p c^4}\frac{1}{1-\exp{\frac{h\nu_p}{k_B T}}} \frac{I^{\mathrm{Ram}}}{45},
\end{equation}
where $\Omega$ is the volume of the structure; $\nu_S$ is the frequency of the scattered light; $c$ is the velocity of light; $h$ is Planck's constant; $k_B$ is Boltzmann's constant; $T$ is temperature; $I^{\mathrm{Ram}}$ is the polarization-averaged Raman-scattering activity written by \cite{10.1103/PhysRevB.54.7830,10.1016/j.diamond.2004.12.007}
\begin{equation}
    I^{\mathrm{Ram}}=45 \alpha^{\prime 2}+7 \beta^{\prime 2}, \label{eq:raman_act} 
\end{equation}
where
\begin{eqnarray}
    &&\alpha^{\prime 2}=\frac{1}{9}\left(\frac{\partial \alpha_{x x}}{\partial Q_p}+\frac{\partial \alpha_{y y}}{\partial Q_p}+\frac{\partial \alpha_{z z}}{\partial Q_p}\right)^2\\
    &&\beta^{\prime 2}=\frac{1}{2}\left[\left(\frac{\partial \alpha_{x x}}{\partial Q_p}-\frac{\partial \alpha_{y y}}{\partial Q_p}\right)^2+\left(\frac{\partial \alpha_{x x}}{\partial Q_p}-\frac{\partial \alpha_{z z}}{\partial Q_p}\right)^2 \right. \nonumber \\ &&\left.+\left(\frac{\partial \alpha_{y y}}{\partial Q_p}-\frac{\partial \alpha_{z z}}{\partial Q_p}\right)^2 + 6 \left(\left(\frac{\partial \alpha_{x y}}{\partial Q_p}\right)^2+ \left(\frac{\partial \alpha_{x z}}{\partial Q_p}\right)^2 \right.\right.\nonumber \\
    &&+\left.\left.\left(\frac{\partial \alpha_{y z}}{\partial Q_p}\right)^2\right)\right]. 
\end{eqnarray}

Here, $\alpha^{\prime}$ is denoted as the derivative of the polarizability tensor, and $\beta^{\prime 2}$ is its anisotropy. Both derivatives are taken with respect to the normal-mode coordinate $Q_p$ of $p$-th phonon mode along the mass-scaled eigenvector obtained from the phonon calculation. This can be expressed as
\begin{equation}
    Q_p = \frac{e_p(k)}{\sqrt{m_k}},
\end{equation}
where $e_p(k)$ is an eigenvector of a phonon mode $p$ at $k$ atom, and $m_k$ is an atomic mass of the $k$ atom. \\
\indent To facilitate the Raman spectrum simulation, the open-source Raman-SC Python script \cite{raman-sc} is implemented to solve the derivatives of $\alpha^{\prime}$ using the centered finite difference scheme by considering static macroscopic dielectric tensor ($\epsilon_{ij}$) calculated based on the density functional perturbation theory. This can be expressed as
\begin{equation}
\frac{\partial \alpha_{ij}}{\partial Q_p} = \frac{V}{4\pi}\frac{\partial \epsilon_{ij}}{\partial Q_p} \approx \frac{V}{4\pi}\frac{\epsilon_{ij}(Q_p + \Delta Q_p) - \epsilon_{ij}(Q_p - \Delta Q_p)}{2 \Delta Q_p}, \label{eq:dielectric}
\end{equation}
where $V$ is the geometry volume and $\Delta Q_p$ is a small displacement.\\
\indent Finally, the Raman activity from Eq.~\eqref{eq:raman_act} can be obtained. Since the simulation of the Raman spectrum is based on zero-kelvin temperature, the simulated spectrum appears as discrete peaks (a $\delta$ function). To enhance visualization, a Lorentzian function with a standard deviation of 5 cm$^{-1}$ has been consistently applied throughout this work.\\

\subsection*{Calculation of Strain}
The strain values analyzed in this study are derived from our previous experiments on a C-based defect \cite{10.1021/acsnano.3c08940}. Through SHG measurements, we observed that local strain causes deviations in the crystal axes, which are no longer separated by the expected 60$^\circ$ characteristic of a hexagonal lattice structure. Using a first-order approximation based on changes in the lattice constant, we quantified the local strain as approximately 1\% per 1$^\circ$ deviation from 60$^\circ$. Consequently, the simulations in this work have applied strain within this range. It is important to note that while the actual strain values may deviate from this simplified approximation, the qualitative trends and conclusions drawn from the simulations are expected to remain robust.

\subsection*{Dynamic Time Warping Calculation}
To analyze the similarity of Raman lineshapes, DTW was selected to be employed in this study. Unlike traditional measures, such as Euclidean distance, which compare spectral intensities point by point, DTW can align two signals non-linearly to account for variations in peak positions or slight frequency shifts, as also implemented by earlier calculation \cite{10.1002/adts.202000039}. This flexibility makes DTW suitable for analyzing Raman spectra, where such shifts may arise due to strain or computational factors. See Supplementary S5 for more details in justification.\\
\indent The DTW distance, denoted as $D_\text{DTW}$, between two Raman spectra $S_1 = \{s_1(i)\}$ and $S_2 = \{s_2(j)\}$, where $i$ and $j$ represent indices of the respective spectra, is computed by minimizing the accumulated cost of aligning $S_1$ and $S_2$ along a warping path $P$. The warping path is defined as a sequence of index pairs
\begin{equation}
P = \{(i_k, j_k) \, | \, k = 1, 2, \dots, K\},  
\end{equation}
where $K$ is the length of the warping path. The goal is to find $P$ that minimizes the total alignment cost
\begin{equation}
D_\text{DTW}(S_1, S_2) = \min_P \sum_{k=1}^K d(s_1(i_k), s_2(j_k)),
\end{equation}
where $d(s_1(i), s_2(j))$ represents the cost of aligning points $s_1(i)$ and $s_2(j)$, often defined as the squared difference
\begin{equation}
d(s_1(i), s_2(j)) = \left(s_1(i) - s_2(j)\right)^2.
\end{equation}
\indent The optimization process is constructed based on the in-house code, benchmarked with the fast DTW code \cite{FDTW}, to compute the minimum accumulated cost based on the following recurrence relation
\begin{eqnarray}
D(i, j) = d(s_1(i), s_2(j)) + \nonumber \\ 
\min \{ D(i-1, j), D(i, j-1), D(i-1, j-1) \},
\end{eqnarray}
where $D(i, j)$ represents the minimum cost of aligning the subsequences $S_1(1:i)$ and $S_2(1:j)$.\\
\indent After computing the accumulated cost matrix, the optimal $P$ can be extracted. Finally, the $D_\text{DTW}$ is determined as the total cost along this path.


\section*{Data availability}
All raw data from this work is available from the authors upon reasonable request. The Raman spectrum for each defect will be uploaded to our online database available under \url{https://h-bn.info} after peer review.

\section*{Notes}
The authors declare no competing financial interest.

\begin{acknowledgments}
This research is part of the Munich Quantum Valley, which is supported by the Bavarian state government with funds from the Hightech Agenda Bayern Plus. This work was funded by the Deutsche Forschungsgemeinschaft (DFG, German Research Foundation) - Projektnummer 445275953, under Germany’s Excellence Strategy - EXC-2111 - 390814868, and as part of the CRC 1375 NOA project C2 (Projektnummer 398816777). T.V. is funded by the Federal Ministry of Education and Research (BMBF) under grant number 13N16292. The authors acknowledge support by the German Space Agency DLR with funds provided by the Federal Ministry for Economic Affairs and Climate Action BMWK under grant number 50WM2165 (QUICK3) and 50RP2200 (QuVeKS). S.S. acknowledges research funding by Mahidol University (Fundamental Fund FF-111/2568: fiscal year 2025 by the National Science Research and Innovation Fund (NSRF)). V.D. acknowledges the SFB-1375 NOA–Project C02 (Projektnummer 398816777). The authors gratefully acknowledge the Gauss Centre for Supercomputing e.V.\ (www.gauss-centre.eu) for funding this project by providing computing time on the GCS Supercomputer SuperMUC-NG at Leibniz Supercomputing Centre (www.lrz.de).
\end{acknowledgments}
\section*{Author contributions}
C.C., V.D., and T.V. designed the study. C.C. performed and analyzed the calculations. V.D. proposed an experimental design. All authors contributed to the discussion and review of the manuscript. T.V. conceived and supervised the project.

\bibliography{main}
\end{document}